\documentclass[11pt,tightenlines,onecolumn,superscriptaddress]{revtex4}

\usepackage{amssymb}
\usepackage{color,graphicx}
\usepackage{graphics}
\usepackage{amsmath}
\usepackage{amsbsy}
\usepackage{amsthm}
\usepackage{bbm}
\usepackage{bm}
\usepackage{epsfig}
\usepackage{lscape}
\usepackage{float}
\usepackage{natbib}

\begin{document}

\title{Are collapse models testable with quantum oscillating systems? \\
The case of neutrinos, kaons, chiral molecules}

\author{M. Bahrami}
\email{mbahrami@ictp.it}
\affiliation{
The Abdus Salam ICTP, Strada Costiera 11, 34151 Trieste, Italy}
\affiliation{Department of Chemistry, K. N. Toosi University of Technology,
1587-4416 Tehran, Iran}

\author{S. Donadi}
\email{sandro.donadi@ts.infn.it}
\affiliation{Department of Physics, University of Trieste, Strada Costiera 11, 34014 Trieste, Italy}
\affiliation{Istituto
Nazionale di Fisica Nucleare, Trieste Section, Via Valerio 2, 34127 Trieste,
Italy}

\author{L. Ferialdi}
\email{ferialdi@ts.infn.it}
\affiliation{Department of Physics, University of Trieste, Strada Costiera 11, 34014 Trieste, Italy}
\affiliation{Istituto
Nazionale di Fisica Nucleare, Trieste Section, Via Valerio 2, 34127 Trieste,
Italy}

\author{A. Bassi}
\email{bassi@ts.infn.it}
\affiliation{Department of Physics, University of Trieste, Strada Costiera 11, 34014 Trieste, Italy}
\affiliation{Istituto
Nazionale di Fisica Nucleare, Trieste Section, Via Valerio 2, 34127 Trieste,
Italy}

\author{C. Curceanu}
\email{Catalina.Curceanu@lnf.infn.it}
\affiliation{Laboratori Nazionali di Frascati dell'INFN, Via E. Fermi 40, 00044 Frascati, Italy}

\author{A. Di Domenico}
\email{antonio.didomenico@roma1.infn.it}
\affiliation{Department of Physics, Sapienza University of Rome,
P.le Aldo Moro 5, 00185 Rome, Italy}
\affiliation{Istituto Nazionale di Fisica Nucleare, Sezione di Roma, P.le Aldo Moro 5, 00185 Rome, Italy}

\author{B. C. Hiesmayr}
\email{Beatrix.Hiesmayr@univie.ac.at}
\affiliation{Masaryk University, Department of Theoretical Physics and Astrophysics, Kotl\'a\v{r}\'ska 2, 61137 Brno, Czech Republic}\affiliation{University of Vienna, Faculty of Physics, Boltzmanngasse 5, 1090 Vienna, Austria}

\begin{abstract}
Collapse models provide a theoretical framework for understanding how classical world emerges from quantum mechanics. Their dynamics preserves (practically) quantum linearity for microscopic systems, while it becomes strongly nonlinear when moving towards macroscopic scale. The conventional approach to test collapse models is to create spatial superpositions of mesoscopic systems and then examine the loss of interference, while environmental noises are engineered carefully. Here we investigate a different approach:  We study systems that naturally oscillate---creating quantum superpositions---and thus represent a natural case-study for testing quantum linearity: neutrinos, neutral mesons, and chiral molecules. We will show how spontaneous collapses affect their oscillatory behavior, and will compare them with environmental decoherence effects. We will show that, contrary to what previously predicted, collapse models cannot be tested with neutrinos. The effect is stronger for neutral mesons, but still beyond experimental reach. Instead, chiral molecules can offer promising candidates for testing collapse models.
\end{abstract}

\maketitle

\section{Introduction}
A great variety  of important physical phenomena can be effectively described in a two-dimensional Hilbert space, when the system's dynamics effectively involves only two relevant states. The most common examples include oscillatory, decaying  and/or relaxation effects in: elementary particles (e.g., neutrino and kaon oscillation~\cite{Beuthe,vogel}), atoms (e.g., Rabi oscillation and spontaneous emission~\cite{agar}), molecules (e.g., tunnelling in double-well potentials, like Ammonia inversion~\cite{chi_book2,bahrami1,bahrami2}), and crystals (e.g., spin relaxation~\cite{spin1,spin2}).

In such systems, oscillations occur because the relevant states are {\it not} eigenstates of the system's Hamiltonian. To be definite, and without loss of generality within the two dimensional formalism, let us take the eigenstates $|\text{+}\rangle$ and $|-\rangle$ of the $\hat{\sigma}_z$ operator as the relevant states, and $\hat{H}_0=\omega_x\,\hat{\sigma}_x/2$ as the Hamiltonian, where $\omega_x$ is the characteristic oscillating frequency (for example, for Ammonia, $\omega_x=24\,$GHz is the inversion frequency). If we start from any eigenstate of $\hat{\sigma}_z$, we observe the coherent oscillation between $|+\rangle$ and $|-\rangle$ with frequency $\omega_x$. In this idealised situation, temporal oscillations remain coherent in time, with a constant amplitude. However, in practice they lose coherence and decay more or less rapidly, because the system is exposed to external noises.  Such environmental effects can be effectively described by Lindblad-type master equations~\cite{decoherence1,decoherence2,decoherence3}.

\begin{figure}
  \centering
    \includegraphics[width=0.7\textwidth]{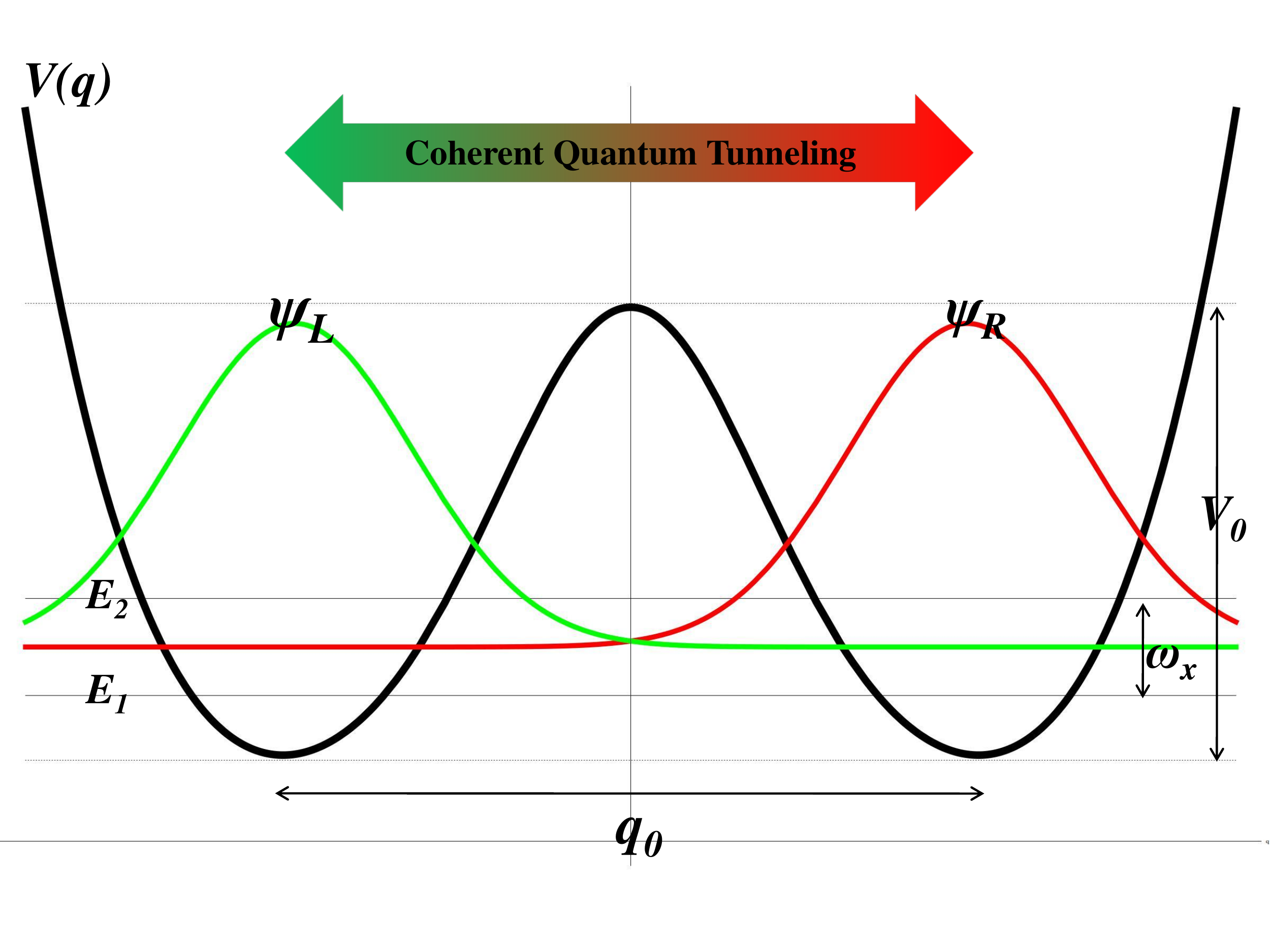}
\caption{{\bf A symmetric double-well potential}. Important parameters are the tunnelling frequency $\omega_x=(E_2-E_1)/\hbar$, the height barrier $V_0$ and the minima separation of $q_0$. The molecular structures are associated with left and right chiral states, $\psi_L$ and $\psi_R$, that are localized in each minima. The tunnelling splitting is manifested as doubling in the spectra of the molecule~\cite{chi_book2}.}
\end{figure}

Oscillations become of great conceptual importance when the two relevant states $|\text{+}\rangle$ and $|-\rangle$  become ``macroscopically" distinct.
This is the typical situation with chiral molecules, as we will see. The observation of oscillations between two such states is directly connected with the highly-debated problem (both theoretically and experimentally) of the quantum-to-classical transition: {\it how linear quantum mechanics copes with macroscopic classical variables}, where ``classical'' implies no superposition~\cite{decoherence1,decoherence2,decoherence3,Bell}. The fundamental question is whether such ``macroscopic oscillations'' persist when the system increases in size (and assuming that environmental sources of noise are kept under control) as predicted by quantum mechanics, or alternatively if they unavoidably decay in time because of {\it intrinsic nonlinear effects} in the dynamics.
This second possibility is predicted by collapse models~\cite{Bell,Grw,Cslmass,Csl,adlerphoto,Pr,Pe1,Pe2,
Di1,Di2,qmupl,Pd,new_phys1,new_phys2,new_phys3,collapse_review2}.

Collapse models have been extensively studied in the literature. There has been also a rapid progress in experimental searches of nonlinear effects predicted by collapse models~\cite{collapse_review2}, in particular by delocalizing large massive objects with matter-wave interferometry and optomechanical techniques~\cite{macro_mol1,macro_mol2,macro_mol3,opto1,opto2,opto3}. In order to motivate further experimental searches of such nonlinear effects, here we follow a different approach by studying how collapse affects naturally oscillating quantum systems. In these cases, it is not necessary to create the superpositions in the laboratory, as they appear spontaneously from the dynamics.

Collapse models add stochastic and nonlinear terms to the Schr\"odinger dynamics, which induce the collapse of the wave function. In the most well-studied collapse models (CSL~\cite{Csl}, QMUPL~\cite{qmupl}), a noise-field is nonlinearly coupled to the spatial degrees of freedom of any massive system, inducing the suppression of spatial coherence. These models are discussed later in the text. Here it suffices to say that, when restricting to a 2D Hilbert space, where the states $|\text{+}\rangle$ and $|-\rangle$ describe two different spatial configurations, then the collapse dynamics takes the form~\cite{bi}:
\begin{eqnarray}\label{eq:SDE_Sch}
d|\psi_t\rangle&=&
\left[-i\frac{\omega_x}{2}\hat{\sigma}_x\,dt
+\sqrt{\lambda}
\left(\hat{\sigma}_z-\langle\hat{\sigma}_z\rangle\right)\,dW_t
-\frac{\lambda}{2}
\left(\hat{\sigma}_z-\langle\hat{\sigma}_z\rangle\right)^2\,dt
\right]|\psi_t\rangle,
\end{eqnarray}
with $W_t$ a standard Wiener process, and $\lambda$ the collapse rate depending on the size of system and the nature of oscillation. 
The last two terms of Eq.~\eqref{eq:SDE_Sch} induce the collapse of the wave function either to $|+\rangle$ or $|-\rangle$, according to the Born probability rule.
In experimental situations, only averages over the noise are relevant. These can be computed from the density matrix $\hat{\rho}_t \equiv {\mathbb E} [|\psi_t \rangle \langle \psi_t| ]$, where ${\mathbb E}[\cdot]$ denotes the stochastic average. It is not difficult to prove that $\hat{\rho}_t$ obeys the following Lindblad-type equation~\cite{bi}:
\begin{equation}
\label{eq:2-state}
\frac{d}{dt}\hat{\rho}_t=
-i \frac{\omega_x}{2}[\hat{\sigma}_x,\hat{\rho}_t]
-\frac{\lambda}{2}[\hat{\sigma}_z,[\hat{\sigma}_z, \hat{\rho}_t]],
\end{equation}
Quantum linearity (manifested by the oscillatory behavior) is well preserved when $\omega_x\gg\lambda$, while nonlinearity (i.e., no quantum superposition) becomes dominant when $\lambda \gg \omega_x$. In this way collapse models provide a quantitative description for the transition from the microscopic quantum world to the macroscopic classical one.

For any given physical system, one has to derive $\lambda$ from the full collapse dynamics, in the same way in which the characteristic frequency $\omega_x$ can be deduced, at least in principle, from the complete Hamiltonian of the system. In the next sections, we will compute $\lambda$ for three different types of oscillatory systems: neutrinos, neutral mesons, and chiral molecules.
We will show that, contrary to what previously predicted~\cite{Ch}, collapse models cannot be tested with neutrinos. The collapse effect is stronger for neutral mesons, but still beyond experimental reach. Instead, chiral molecules offer promising candidates for testing collapse models.

Eq.~\eqref{eq:2-state} has the same form as that describing an oscillatory system under environmental noises~\cite{decoherence1,decoherence2,decoherence3}. This means that, in analysing the effect of collapse models on oscillating systems, one has to consider also environmental effects, which tend to mask the collapse effects, by damping oscillations in a similar way. In each case, we will compare predictions of collapse models with decoherence effects.


\section*{Results}
We present the analysis of how collapse models modify the oscillatory behavior of neutrinos, neutral mesons and chiral molecule. We will use the mass-proportional Continuous Spontaneous Localization (CSL) model~\cite{collapse_review2}; details of CSL dynamics are explained in Methods section. The representation of the collapse dynamics in the position-basis predicts the following collapse rate for the off-diagonal elements of the density matrix of a generic system consisting of $N$ nucleons:
\begin{equation}\label{eq:csl-chi}
    \lambda_{\tiny {\text{CSL}}}=\frac{\Lambda}{2}\sum_{i,j=1}^{N}
    \left[F(\textbf{x}'_i-\textbf{x}'_j)+F(\textbf{x}''_i-\textbf{x}''_j)
    -2F(\textbf{x}'_i-\textbf{x}''_j)\right],
\end{equation}
where $\Lambda \simeq 10^{-9}\,\text{Hz}$ (see Methods), $N$ is the number of nucleons in the spatial superposition,
$F(\textbf{r}) =  \exp[-\textbf{r}^2/4r_C^2]$ with $r_C = 10^{-5}\text{cm}$, and $\{\textbf{x}'_i\},\,\{\textbf{x}''_i\}$ are distinct positions of nucleons in spatial superposition.

The connection between the full characterization of $\lambda_{\tiny {\text{CSL}}}$, given by Eq.~\eqref{eq:csl-chi}, with the two-dimensional one given by Eq.~\eqref{eq:2-state} is not always straightforward. It depends on the system under study, and sometimes needs careful analysis and lengthy calculations, as we will show.

In applying collapse models to experiments, one has always to take decoherence effects into account, as they produce apparently similar effects. For neutrinos and chiral molecules, collisions are the dominant source of decoherence. Using collisional decoherence theory~\cite{decoherence1,decoherence2,decoherence3}, we exploit the phenomenological formula: $\lambda_{\text{\tiny DEC}}\sim n\,v\,\sigma_{\text{\tiny DEC}}$ (with $v$ the relative velocity, $n$ the density of bath particles, and $\sigma_{\text{\tiny DEC}}$ the decoherence scattering cross section) in order to estimate the decoherence rate. For mesons, we provide upper bounds on $\lambda_{\text{\tiny DEC}}$ using available experimental data.

\subsection*{Neutrino Oscillation}
\noindent {\bf Effective description of neutrino oscillations.} The flavour eigenstates of neutrinos $\left|\nu_{\alpha}\right\rangle$ (with $\alpha=e,\mu,\tau$ for electronic, muonic and tauonic neutrinos) are linear combinations of mass eigenstates:
$\left|\nu_{\alpha}\right\rangle =\sum_{j=1}^{3}\text{U}_{\alpha j}\left|\nu_{j}\right\rangle$,
with $\hat{\text{U}}$ the unitary mixing matrix.
Therefore, for a neutrino in an initial flavour eigenstate, the transition probability between different flavour eigenstates shows an oscillatory behaviour in the course of time~\cite{Beuthe,vogel}.
This oscillation may be damped either by environmental interactions, or by nonlinearities in the dynamics such as those predicted by collapse models.
Neutrinos are the lightest massive particles, therefore it seems unlikely that they show any spontaneous collapse effect. However, they can travel very long distances through space, and there could be enough time  during the flight, for collapse effects to build up appreciably. Therefore, it is not clear beforehand whether neutrinos can play any role in testing spontaneous collapses.

\noindent {\bf The collapse rate in neutrino oscillations.}
The effect of collapse models on neutrino oscillation was first elaborated by Christian~\cite{Ch}, using the Di\'{o}si-Penrose (DP) gravity-induced collapse model~\cite{Di1,Di2,Pd}. Gravity is fundamentally nonlinear, therefore when properly taken into account, it induces a nonlinear modification of the Schr\"odinger equation.
According to the analysis of Christian (see also~\cite{neuCSL}), the predicted magnitude of the oscillation damping ($\lambda t$) is between $\sim 10^{-2}$ and $\sim 1$ for cosmogenic neutrinos. This value is strong enough to be tested with high-precision techniques.
However, this strong predicted effect is questionable, for the following reason. For point-like constituents, like neutrinos,  gravitational self-energy diverges, implying a divergence in DP model. To avoid this problem, Di\'{o}si~\cite{Di1,Di2} originally introduced a cutoff for small lengths, equal to the nuclear size. However, Ghirardi, Grassi and Rimini~\cite{ggr} showed that a much larger cutoff ($\sim10^{-7}\,$m) is needed,  in order for the model to be consistent with known experimental data. On the other hand, the effective size of the neutrinos, as introduced by Christian, is $\sim10^{-30}\,$m, well beyond any reasonable cutoff. Therefore the result cannot be trusted.

We compute the collapse effect on neutrino oscillations using the CSL model, which is free from the divergences contained in the DP model (apart from standard quantum field theoretical ones, which can be treated with usual renormalization techniques).
The dynamics of neutrino oscillation is phenomenologically described in a 3D--Hilbert space of flavour, while the CSL collapse occurs in space. Therefore, the major task is to link the spatio-temporal description of Eq.~\eqref{eq:csl_chi} to the 3D--dynamics neutrino oscillations. Since the CSL model is a field theoretical model, this can be properly done by resorting standard quantum field theoretical techniques. Treating the noise as perturbation, for the collapse rate we obtain~\cite{neuCSL}:
\begin{eqnarray}\label{eq:xi}
\lambda_{jk} &=& \frac{\Lambda}{2m_{0}^{2}c^{4}}
\left(\frac{m_{j}^{2}c^{4}}{E_{i}^{\left(j\right)}}
-\frac{m_{k}^{2}c^{4}}{E_{i}^{\left(k\right)}}\right)^{2}\,,
\end{eqnarray}
where $m_0=1\,$amu, $m_j$ are the eigenvalues of mass-eigenstates, and $E_i^{(j)}=\sqrt{p_i^2c^2+m_j^2c^4}$ with $p_i$ the momentum.
In the relativistic regime, as appropriate for neutrinos, one has $E_i^{(j)}\simeq p_ic$. By taking the largest mass difference in Eq.~(\ref{eq:xi}), one finds the following upper bound:
\begin{equation}\label{eq:xi2}
\lambda_{ij}\,t\leq 7\times 10^{-36}\frac{t/t_0}{(E/E_0)^2},
\qquad\qquad
t_0=1\,\text{s}, \qquad E_0=1\,\text{eV},
\end{equation}
where the energy ($E$) and the time of flight ($t$) of the neutrinos depend on the type of neutrinos under study. In Table.~I, this damping factor has been computed for neutrinos originating from three different sources.
The CSL collapse effect is very tiny and non detectable with present-day technology, the reason being that neutrinos are too light, although they can travel long distances.

\noindent {\bf Decoherence effects in neutrinos.} We also analyze decoherence effects on neutrino oscillations due to the scattering with particles (mainly leptons), during their flight through space. The experimental value of the relevant scattering cross section, $\sigma_\text{\tiny DEC}$, are known in the literature~\cite{sigmanue1,sigmanue2}. The average density of electrons in outer space and in the atmosphere are respectively $n_e^{\text{\tiny OUT}}\sim1/\mathrm{m}^3$ and
$n_e^{\text{\tiny ATM}}\sim2\times10^{26}/\mathrm{m}^3$,
while the average density of neutrinos is about $n_{\nu}\sim10^8/\mathrm{m}^3$ everywhere (electrons and neutrinos are the two main sources of decoherence~\cite{neuCSL}).
Assuming the neutrino velocity $v$ equal to the velocity of light in vacuum, we get:
\begin{eqnarray}
\lambda_{\text{\tiny DEC}}^{\text{\tiny OUT}} \sim
\frac{10^{-43}E}{E_0}\,\text{Hz}\,,\quad
\lambda_{\text{\tiny DEC}}^{\text{\tiny ATM}}\sim\frac{10^{-20}E}{E_0}
\,\text{Hz},
\end{eqnarray}
with $\lambda_{\text{\tiny DEC}}^{\text{\tiny OUT[ATM]}}$ the decoherence rate in the out-space [atmosphere]. Neutrinos travel through the atmosphere within $\sim10^{-4}\,$s, the remaining time being spent in traveling through outer space. Taking both contributions from atmosphere and outer space into account, and using data listed in Table.~I, the decoherence damping factor for cosmogenic neutrinos (CN) turns out to be: $\lambda_{\text{\tiny CN}} t \sim10^{-5}$. For solar neutrinos (SN) instead, one gets: $\lambda_{\text{\tiny SN}} t  \sim10^{-18}$,
which is hardly detectable, in agreement with well-known experimental results~\cite{decoexp,vogel}.

This analysis shows that, since environmental decoherence on neutrino oscillations is much stronger than the CSL collapse effect (and comparable with that---overestimated---predicted by Christian~\cite{Ch}), even if technology were able in principle to discriminate collapse effects on neutrino oscillations, these effects would be masked by unavoidable decoherence effects.

\subsection*{Neutral Mesons}
\noindent {\bf Effective description of neutral mesons oscillations.} As a second example of oscillating quantum systems, we consider neutral mesons.
Differently from neutrinos, they offer the advantage that decoherence effects can be kept low, since they are produced in a very controlled environment.

A meson consists of a quark and an antiquark. For example there exists the
neutral K-meson ($K^0$, made of $\bar s$ and $d$; or $\bar K^0$, made of $s$ and $\bar d$) or the neutral B-meson system ($B^0$, made of $\bar b$ and $d$; or $\bar B^0$, made of $b$ and $\bar d$). The phenomenology of these oscillating and decaying systems is usually described by a $2\times 2$ non-Hermitian Hamiltonian whose stationary states are the mass eigenstates~\cite{kaonCSL}: $\hat{H}_{\text{\tiny eff}}|M_{1,2}\rangle=(m_{1,2}-\frac{i}{2}\Gamma_{1,2})|M_{1,2}\rangle$ where $m_{1,2}$ and $\Gamma_{1,2}$ are masses and decay widths.
The mass eigenstates are related to the flavour eigenstates via $|M_{1,2}\rangle=[|M^0\rangle\pm|\bar M^0\rangle]/\sqrt{2}$, if and only if we assume $\mathcal{CPT}$ conservation and neglect $\mathcal{CP}$ violation. We can safely assume such a linear combination because the $\mathcal{CP}$ violation is a very small effect in our case.

\noindent {\bf The collapse rate for neutral mesons' oscillations.}
In computing the predictions of collapse models for the oscillations in neutral mesons, we follow the same approach as the one we did for neutrino oscillations. We perform the computation by expanding the CSL dynamics to the first significant perturbative order, in order to find the dominant effect. The calculation is long but straightforward, and is fully reported in~\cite{kaonCSL}. The final result for the collapse rate is:
\begin{eqnarray}\label{decovalue}
\lambda_{\text{\tiny CSL}} =\frac{\Lambda\;(m_2-m_1)^2}{2\, m_0^2}.
\end{eqnarray}
We list the damping rates in Table.~I for distinct mesons, using the experimental values given in Ref.~\cite{ParticleDataBook}.
With not much surprise, the obtained values are much larger than those for neutrinos.

\noindent {\bf The decoherence rate in mesons.}
Environmental decoherence effects have been investigated~\cite{BeatrixZeta3} and compared with experimental data~\cite{DiDomenico1,DiDomenico2,DiDomenico3,Appollo,CPLEAR}.
Bounds from experimental data of the CPLEAR experiment~\cite{CPLEAR} and to the more refined data of the KLOE experiment of the DAPHNE collider~\cite{DiDomenico1,DiDomenico2,DiDomenico3} were obtained in terms of a phenomenological time-independent parameter $\zeta$~\cite{BeatrixZeta3}. This parameter, first introduced by Schr\"odinger, quantifies the spontaneous factorization of an initially  entangled  wave function in a chosen basis. The best value, obtained by measuring $2$-pion final states, is: $\zeta=0.003\pm0.018_\textrm{stat}\pm 0.006_\textrm{syst}$. Since this is a time averaged quantity, we can use it only for small times, when $\zeta\approx \lambda_{\text{\tiny deco}} t$. From that we may deduce an upper bound on the decoherence rate which is about $8 \times 10^7$Hz with $\%90$ confidence level. Thus, comparing this value with those in Table.~I, we see that collapse models are not directly measurable for strangeness oscillations; for other types of mesons similar considerations hold. To test collapse models for mesons one has to find observables being more sensitive to the CLS effect.


\subsection*{Chiral Molecules}
\noindent {\bf Effective description of chiral molecules.} Another very relevant example of oscillating quantum system is given by chiral molecules, in which case $|\text{+}\rangle,\,|-\rangle$ represents two configurations with different macroscopic properties, e.g., optical activity. The classical example is Ammonia inversion phenomenon~\cite{chi_book2}. In general, non-rigid molecules and molecular complexes have at least two stable configurations that can be transformed to each other by
a large-amplitude vibration~\cite{chi_book2}.  In the zero-th order approximation, this vibration can be described by the motion of a particle of effective mass $\mu$ in a  double-well potential $V(q)$, where $q$ is a generalized large-amplitude coordinate. The minima of the wells are positioned at $q =\pm q_0/2$, separated by a barrier $V_0$ (see Fig.~1). Molecular configurations are described by localized states (say ``chiral" states) at each minima. The tunnelling through the height barrier leads to measurable level splittings in the molecular spectra, which has been observed for a large variety of non-rigid molecules and molecular complexes~\cite{chi_book2,resol1,resol2}.

In the limit $V_0 \gg \omega_0 \gg k_BT$ (where $T$ is temperature, $k_B$ is Boltzmann constant, and $\omega_0 = [V''(\pm q_0/2)/\mu]^{1/2}$ is the small-amplitude vibration in either well), the state of the molecule is effectively confined in the two-dimensional Hilbert space spanned by two chiral states~\cite{bahrami1,bahrami2}.
Thus, the Hamiltonian becomes $\hat{H}_0=\omega_x\hat{\sigma}_x/2$ with $\omega_x$ the level splitting due to the tunnelling.

\noindent {\bf The collapse rate in chiral molecules.}
We consider superpositions of chiral states as spatial superpositions of an atom or group of atoms between two distinct molecular configurations. So differently from the case of neutrinos and neutral mesons, we can immediately derive the collapse rate from Eq.~\eqref{eq:csl-chi}.
Typical non-rigid molecules are within a range of size $1-100\,$\AA. This implies that chiral coherence is distributed over the region whose dimension is much smaller than $r_C=10^3\,$\AA. We can then expand $F(\mathbf{r})$ to the leading order of $\mathbf{r}$ in Eq.~\eqref{eq:csl-chi}, and we obtain:
\begin{equation}
\label{eq:csl_chi}
    \lambda_{\text{\tiny CSL}}\simeq
    \frac{\Lambda}{4 r_C^2} \left(\sum_{i=1}^{n}m_i(\textbf{x}^{L}_i-\textbf{x}^{R}_i)\right)^2,
\end{equation}
where $m_i$ is the mass (in amu) of $i$-th atom,
$n$ is the number of atoms in the spatial superposition (e.g., for Ammonia, we have three Hydrogen atoms in superposition; $n=3$), and ${\bf x}^{L}_i$ and ${\bf x}^{R}_i$ are positions of $i$-th atom in the two chiral conformations where the origin is the chirality center (e.g., for Ammonia, they are positions respect to the Nitrogen atom).
The mere knowledge of positions of atoms in chiral structures of the molecule is enough to compute the collapse rate using Eq.\eqref{eq:csl_chi}. However, when data about the effective mass $\mu$ and the minima separation $q_0$ of the double-well potential is available, then one can simply use the following simpler formula for the collapse rate:
\begin{eqnarray}
\label{eq:csl_chi2}
\lambda_{\text{\tiny CSL}}&\approx&\frac{\Lambda}{4r_C^2}\left(\mu\,q_0\right)^2,
\end{eqnarray}
with $\mu$ the effective mass (in amu) moving in the double-well potential. For example, $\mu=m_{\text{\tiny N}}(3m_{\text{\tiny H}})/(m_{\text{\tiny N}}+3m_{\text{\tiny H}})\approx3\,$amu and $q_0=0.8$\AA \; in the case of Ammonia~\cite{chi_book2}.

We apply Eq.~\eqref{eq:csl_chi} to compute the collapse rates for some pyramidal chiral sulfoxides~\cite{so}. The results are shown in Table.~I. For each sulfoxide, we obtain the enantiomeric equilibrium structures with DFT (B3LYP), using a minimum basis set by Firefly program~\cite{firefly}. As expected, the collapse rates are by many orders of magnitude stronger than those of neutrinos and mesons.

\noindent {\bf Decoherence rate in chiral molecules.}
If we model the chiral coherence as the spatial superposition of a quantum Brownian particle of effective mass $\mu$ over the distance $q_0$ (see next section for more detail), then we can use the linearized quantum Brownian dynamics to compute $\sigma_{\text{\tiny DEC}}$~\cite{bahrami1,bahrami2}. In this way we can compute the dominant contribution to $\lambda_{\text{\tiny DEC}}$ by using  Eqs.~(3) and~(13) of Ref.~\cite{bahrami2}. We consider the London dispersion potential for collisions. Then, for the density of background gas about $\sim10^{10}$particles/m$^3$ (the conventional ultra-high vacuum) and the background temperature of $T\simeq300\,$K, we obtain: $\lambda_{\text{\tiny DEC}}\sim10^{-6}$--$10^{-4}$Hz.
Considering the cryogenic vacuum where $n\sim10^{5}\,$particles/m$^3$~\cite{vacuum}, then we get: $\lambda_{\text{\tiny DEC}}\sim10^{-11}$--$10^{-9}$Hz.
Accordingly, for chiral molecules decoherence can be practically reduced to a negligible level compared to collapse effects (see Table.~I), thus quantum nonlinearities can be in principle tested using chiral coherence.

\subsection*{Estimates of bounds on $\Lambda$}


We showed that, contrary to the cases of neutrinos and mesons, spontaneous collapse effects (quantified by $\lambda$) can be in principle tested with chiral molecules because environmental effects can be  controlled in such a way that the decoherence becomes negligible. Of course, the great challenge is to find a feasible experimental scheme. We leave the question to future research. In the meantime, one can follow a different strategy and use the spectroscopic data of tunnelling splittings to introduce upper bounds on $\Lambda$. Here, we discuss this strategy.

According to Eq.~\eqref{eq:2-state}
the dipole moment $\langle\hat{\sigma}_z(t)\rangle = \text{Tr}[\rho(t) \hat{\sigma}_z]$  shows no oscillation when $\lambda \geq \omega_x$.
If this is the case, then the spectra of the molecule should show no tunnelling splitting.
Accordingly, the experimental observation of a tunnelling frequency $\omega_x$ implies that $\lambda < \omega_x $. This places an upper bound on the collapse parameter $\Lambda$, which according to Eq.~\eqref{eq:csl_chi2} can be written as follows:
\begin{eqnarray}
\label{eq:bound}
\Lambda < \left(\frac{2r_C}{\mu\,q_0}\right)^2\omega_x.
\end{eqnarray}
The smaller the observed tunnelling frequency, the stronger the bound (see Table.~II).
To our best knowledge, the smallest molecular tunnelling splitting that has been observed is of the order of a few Hertzs for Ru-D$_2$ complex with NMR spectroscopy~\cite{tunnel2}, where $\mu=2\,$amu and $q_0\sim1-2\,$\AA. Accordingly, we get
$\Lambda<10^{5}\,$Hz, which is $10^{14}$ times larger than the standard CSL value.
This should be compared with one of the best available experimental bounds on the collapse rate: $\Lambda<10^{-5}\,$Hz, which is obtained by quantum interfere of massive objects with a mass of $7\times10^3\,$amu, in matter-wave interferometry experiment~\cite{macro_mol2}. So, also this bound is very week, but could be significantly improved.

According to Eq.~\eqref{eq:bound}, in order to obtain stronger bound on $\Lambda$, one should move toward smaller tunnelling frequencies (i.e., smaller $\omega_x$) or larger effective sizes (i.e., larger $\mu$ and $q_0$). In molecular systems, the effective size that can be simply described by a double-well potential, is limited in the ranges $q_0\sim1-10\,$\AA$\;$ and $\mu\sim1-100\,$amu~\cite{chi_book2}. According to Eq.~\eqref{eq:bound},
we get: $\Lambda < \alpha\,\omega_x$, with $\alpha$ varying in the range $4-4\times10^6$.
Therefore, among possible strategies for testing collapse effects in molecular systems, the observation of smaller tunnelling frequencies is the most flexible.

This strategy becomes even more promising if we consider recent progresses in high-resolution spectroscopic methods~\cite{resol1,resol2}.
As we discussed before, $\omega_x$ is manifested as level splittings in 
rotational-vibrational spectra, where molecular modes cover the typical frequency range of $\omega=10^{9}-10^{14}\,$Hz (from microwave to UV~\cite{chi_book2,resol1,resol2}).
Thus, if we use a spectroscopic method with resolution $R=\omega_x/\omega$, we find:
$\Lambda < \beta\,R$, where $\beta$ varies in the range $4\times10^{9}-4\times10^{20}\,$Hz.
So, with a relative resolution $R\sim10^{-14}$, which is in the range of available highest resolution spectroscopy techniques~\cite{resol1,resol2}, one can reach a bound for $\Lambda$, comparable with that obtained in matter wave interferometry. With better resolutions, we can set stronger bounds.
Molecules of the form Y-X-X-Y may serve as candidate molecules where their torsional internal rotation can be simply described by a double-well potential~\cite{chi_book2,resol1,resol2}, with a very tiny tunnelling splitting when Y is a heavy atom (e.g., for Cl$_2$O$_2$, we have the theoretical value of $\omega_x\sim10^{-11}\,$Hz~\cite{resol1,resol2}).

\section*{Discussion}
We computed the predictions of the Continuous Spontaneous Localization (CLS) collapse model for the damping of the oscillatory behavior of three distinct naturally oscillating quantum systems: neutrinos, mesons and chiral molecules. The numerical results are summarized in Table.~I. We also analysed the main decoherence effects on these systems and compared them with the predictions of the CSL collapse model. The values we obtained for the  collapse rates for the first two types of systems are much smaller than the main decoherence rates; consequently, possible violations of the superposition principle cannot be directly observed in these oscillatory systems.
Chiral molecules are better candidates. We suggest a new type of experiment with chiral molecules, which can serve as a test of quantum linearity, and which can possibly put stronger upper bounds on the collapse parameters, than those already available from the literature.

Our formulation of chiral molecules also includes any system whose effective dynamics is described by a double-well potential. A very promising line of research is the study of systems that can be artificially prepared in a double-well potential where its parameters ($V_0$, $\mu$ and $q_0$) are  adjustable at will. Then, by tuning them in proper ranges, e.g. larger $\mu$, one can hope to set further bounds on the collapse parameters.

\begin{table}
\begin{tabular}[b]{p{5cm}p{3cm}p{3.5cm}r}
\hline
\multicolumn{4}{c}{\textbf{NEUTRINOS}} \\
\hline
\textbf{Types of neutrinos} & \textbf{Energy} (eV) & \textbf{Time of Flight} (s) & \textbf{CSL damping} ($\lambda_{ij}t$) \\
Cosmogenic neutrino
& $10^{19}$ & $3\times10^{18}$ & $2\times10^{-55}$ \\
Solar neutrino
& $10^{6}$ & $5\times10^{2}$ & $4\times10^{-45}$ \\
Laboratory neutrino
& $10^{10}$ & $2\times10^{-2}$ & $2\times10^{-57}$ \\
\textbf{Decoherence effect}
 & \multicolumn{3}{r}{$\lambda_{\text{\tiny DEC}}t\sim10^{-18}-10^{-5}$} \\
\hline
\multicolumn{4}{c}{\textbf{NEUTRAL MESONS}} \\
\hline
 {\bf Types of mesons} & \multicolumn{3}{r}{{\bf CSL collapse rate} $\lambda_{\text{\tiny CSL}}\,$(Hz)} \\
K-meson
 & \multicolumn{3}{r}{$1.5\times10^{-38}$} \\
B-meson
 & \multicolumn{3}{r}{$1.4\times10^{-34}$} \\
B$_\text{s}$-meson
 & \multicolumn{3}{r}{$1.7\times10^{-31}$} \\
D-meson
 & \multicolumn{3}{r}{$3.2\times10^{-37}$} \\
\textbf{Decoherence effect}
 & \multicolumn{3}{r}{$\lambda_{\text{\tiny DEC}}\leq 8\times10^7$} \\
\hline
\multicolumn{4}{c}{\textbf{CHIRAL MOLECULES}} \\
\hline
 {\bf Type of molecule} & 
 & \multicolumn{2}{r}{{\bf CSL collapse rate} $\lambda_{\text{\tiny CSL}}\,$(Hz)} \\
SOCH$_3$(p-CH$_3$C$_6$H$_4$)
 & 
 & \multicolumn{2}{r}{$6.3\times10^{-10}$} \\
SOCH$_3($C$_6$H$_5$)
 & 
 & \multicolumn{2}{r}{$7.9\times10^{-10}$} \\
SOCH$_3$(CH$_2$CH$_{2}$-$\alpha$-C$_{10}$H$_7$)
 & 
 & \multicolumn{2}{r}{$2.5\times10^{-9}$} \\
SOCH$_3$(1-pyrenyl)
 & 
 & \multicolumn{2}{r}{$5\times10^{-9}$} \\
\textbf{Decoherence effect}
&\multicolumn{3}{r}{$\lambda_{\text{\tiny DEC}}\sim10^{-11}-10^{-9}\,$}\\
\hline
\end{tabular}
\caption{{\it Theoretical values of CSL collapse rate and decoherence rate for neutrinos, mesons and chiral molecules}. By moving from elementary particle scale to molecular scale, the collapse rate $\lambda_{\text{\tiny CSL}}$ increases significantly. The decoherence hides collapse effects in neutrinos and mesons, but it can be reduced at a negligible level compared with collapse rates of chiral molecules. Results show that quantum linearity can be in principle tested using chiral molecules. However, engineering a proper experiment is not straightforward.}
\end{table}

\begin{table}
\begin{center}
\begin{tabular}[b]{p{6cm}p{5cm}c}
\hline
{\bf Molecule} & \textbf{Upper bound on} $\Lambda$ & \textbf{tunnelling splitting} $\omega_x$
\\
\hline
Ammonia~\cite{chi_book2} & $\Lambda<10^{16}\,$Hz & $24\times10^9\,$Hz
\\
Carboxylic acid dimers~\cite{dimer} & $\Lambda<10^{8}\,$Hz & $\sim10^3\,$Hz
\\
Ru-D$_2$ complex~\cite{tunnel2} & $\Lambda<10^{5}\,$Hz & $1-100\,$Hz
\\
High resolution spectroscopy & $\Lambda < 10^{-5}\,$Hz (proposal)& -
\\
{\bf Matter-wave interference}~\cite{macro_mol2} & $\Lambda<10^{-5}\,$Hz & -
\\
{\bf Adler's CSL value}~\cite{adlerphoto} & $\Lambda\sim10^{-9}\,$Hz & -
\\\hline
\end{tabular}
\caption{{\it Current bounds on the collapse constant $\Lambda$, coming from observation of tunnelling}.
We used available spectroscopic data about tunnelling splittings (see main text). These bounds should be compared with the best experimental bound on $\Lambda$, which is obtained by wave-matter interferometry of molecules with mass $m=7 \times 10^3\,$amu~\cite{macro_mol2}. Using a molecular spectroscopic technique with relative resolution of $R\leq10^{-14}$, it is in principle possible to introduce bounds on $\Lambda$, which could compete with those coming from matter-wave interferometry.}
\end{center}
\end{table}

\section*{Methods}
\noindent {\bf CSL model}.
We consider the most commonly used collapse models in the literature: the mass proportional Continuous Spontaneous Localization (CSL) model~\cite{Cslmass}. The CSL dynamics is:
\begin{equation} \label{eq:dffgdg}
d\psi_t =
\left[-\frac{i}{\hbar}\hat{H}\,dt
+\frac{\sqrt{\gamma}}{m_0}\int d\mathbf{r}\,
\left(\hat{M}(\mathbf{r})-\langle\hat{M}(\mathbf{r})\rangle\right)\,dW_t(\mathbf{r})
-\frac{\gamma}{2m^2_0}\int d\mathbf{r}\,
\left(\hat{M}(\mathbf{r})-\langle\hat{M}(\mathbf{r})\rangle\right)^2\,dt
\right]\psi_t,
\end{equation}
with $\hat{H}$ the standard quantum Hamiltonian, $\langle\hat{M}(\mathbf{r})\rangle \equiv \langle \psi_t | \hat{M}(\mathbf{r}) | \psi_t \rangle$ the standard quantum average (here is where nonlinearity enters the equation),
$m_0=1\,$amu,
$\gamma>0$ the strength of the collapse process, which is a new phenomenological constant of the model,
$W_t(\mathbf{r})$ an ensemble of independent Wiener processes, one for each point in space, and:
\begin{equation}
\hat{M}({\bf r})=\sum_j m_{j}\int
d\mathbf{r}'\,G\left(\mathbf{r'-r}\right)
\hat{a}_{j}^{\dagger}\left(\mathbf{r}'\right)\hat{a}_{j}\left(\mathbf{r'}\right),
\end{equation}
where $\hat{a}_{j}\left(\mathbf{r}\right)$ is the annihilation operator of a particle of type $j$ at position $\mathbf{r}$, and:
\begin{equation}
G({\bf r}) =
\frac{1}{(\sqrt{2\pi}r_{C})^{3}}\exp(-\mathbf{r}^{2}/2r_{C}^{2}),
\end{equation}
with $r_C$ the correlation length, the other new phenomenological constant of the model.
After averaging over all possible realizations of the stochastic processes,
the dynamics for the density matrix is given by:
\begin{equation}
\label{eq:csl-mass}
\frac{d}{dt}\hat{\rho}=
-\frac{i}{\hbar}[\hat{H},\hat{\rho}]
-\frac{\gamma}{2m^2_0}\int d\mathbf{r}\,
[\hat{M}(\mathbf{r}),[\hat{M}(\mathbf{r}), \hat{\rho}]],
\end{equation}
where the second term on the right hand side is the collapse term. It tells that superpositions of states which are closer than $r_C$ are efficiently localized, while superpositions of terms which are further separated are suppressed, with a rate proportional to $\gamma$ and to the size of the system~\cite{adlerphoto}. 

\noindent {\bf Values of collapse parameters}.
The value of the correlation length is commonly fixed to $r_C \simeq 10^{-5}\text{cm}$~\cite{Csl}. For the collapse strength $\gamma$, two values have been proposed in the literature. Ghirardi, Pearle and Rimini~\cite{Csl} set  $\gamma \simeq 10^{-30}\text{cm}^{3}\text{s}^{-1}$, while Adler~\cite{adlerphoto} sets  $\gamma \simeq 10^{-22}\text{cm}^{3}\text{s}^{-1}$.
These values are in agreement with all known experimental data. Much larger values are ruled out because the collapse would become so strong to be detectable also for isolated microscopic systems, contrary to experimental evidence. Much smaller values are also ruled out, because in such cases the collapse would become so weak that the localization of the wave function of macroscopic objects would not be guaranteed anymore. Without this, collapse models would lose their interest. In our analysis, we consider the strongest value of $\gamma$ suggested by Adler.
By defining the collapse rate as: $\Lambda=\gamma/(8\pi^{3/2}r_C^3)$, we get: $\Lambda \simeq10^{-9}\,$Hz. This is the numerical value we used in the text.

\section*{Acknowledgements}
All authors wish to thank the COST Action MP1006 ``Fundamental Problems in Quantum Physics".
BCH acknowledges support from the project Austrian Science Fund: FWF-P21947N16. The Project is funded by the SoMoPro programme. Research leading to these results has received a financial contribution from the European Community within the Seventh Framework Programme (FP/2007-2013) under Grant Agreement No. 229603. The research is also co-financed by the South Moravian Region.
SD, LF, and AB acknowledges partial support from INFN.
AB Acknowledges support also from the EU project NANOQUESTFIT and the John Templeton Foundation project `Quantum
Physics and the Nature of Reality'.
MB acknowledges hospitality from The Abdus Salam ICTP, where this work was carried out.
AB and MB thank  Dr. M. Nahali of SISSA, Italy, for discussions on chiral molecules.
MB thanks Prof. Dr. Gerd Buntkowsky of Technische Universit\"{a}t Darmstadt and
Prof. Dr. H. H. Limbach of Freie Universit\"{a}t Berlin, for data on Ru-D$_2$ complexes.



\begin{thebibliography}{1}

\bibitem{Beuthe}
Beuthe, M. Oscillations of neutrinos and mesons in quantum field theory. {\it Phys. Rep.} {\bf 375}, 105-218 (2003).

\bibitem{vogel}
McKeown, R. D., Vogel, P. Neutrino masses and oscillations: triumphs and challenges. {\it Phys. Rep.} {\bf 394}, 315-356 (2004).

\bibitem{agar}
Agarwal, G. S. {\it Quantum Statistical Theories of Spontaneous Emission and their
Relation to other Approaches}. Springer Tracts Mod. Physics 70, Springer, Berlin (1974).

\bibitem{chi_book2}
Herzberg, G. \textit{Molecular Spectra and Molecular Structure. Electronic Spectra and Electronic Structure of Polyatomic Molecules}. Krieger: Malabar, FL (1991) Vol. II, Sec.II,5(d).

\bibitem{bahrami1}
Bahrami, M., Bassi, A. tunnelling properties of nonplanar molecules in a gas medium. {\it Phys. Rev. A.} \textbf{84}, 062115-8 (2011).
\bibitem{bahrami2}
Bahrami, M., Shafiee, A., Bassi, A. Decoherence effects on superpositions of chiral states in a chiral molecule. {\it Phys. Chem. Chem. Phys.} \textbf{14}, 9214-8 (2012).


\bibitem{spin1}
Ardavan, A. {\it et. al.}, Will Spin-Relaxation Times in Molecular Magnets Permit Quantum Information Processing?. {\it Phys. Rev. Lett.} 98, 057201-4 (2007)
\bibitem{spin2}
Zutic, I., Fabian, J. and Das Sarma, S. Spintronics: Fundamentals and applications. {\it Rev. Mod. Phys.} {\bf 76}, 323-410 (2004).

\bibitem{decoherence1}
Breuer, H.-P., Petruccione, F. {\it The Theory Of Open Quantum Systems}. (Oxford Univ. Press, 2002).
\bibitem{decoherence2} Joos, E., Zeh, H. D., Kiefer, C., and Giulini, D. J. W. {\it Decoherence And The Appearance Of A Classical World In Quantum Theory}. Springer (2003).
\bibitem{decoherence3} Schlosshauer, M. A.  {\it Decoherence And The Quantum-To-Classical Transition}, Springer (2007).

\bibitem{Bell}
Bell, J. S.  {\it Speakable and Unspeakable in Quantum Mechanics}, Cambridge University Press (1986).

\bibitem{Pe1}
Pearle, P. Reduction of the state vector by a nonlinear Schr\"{o}dinger equation. {\it Phys. Rev. D} {\bf 13}, 857-868 (1976).
\bibitem{Pe2}
Pearle, P. Combining stochastic dynamical state-vector reduction with spontaneous localization. {\it Phys. Rev. A} {\bf 39}, 2277-89 (1989).

\bibitem{Grw}
Ghirardi, G. C., Rimini, A., and Weber, T. Unified dynamics for microscopic and macroscopic systems. {\it Phys. Rev. D} {\bf 34}, 470-491 (1986).

\bibitem{Csl}
Ghirardi, G.C., Pearle, P., and Rimini, A.  Markov processes in Hilbert space and continuous spontaneous localization of systems of identical particles. {\it Phys. Rev. A} {\bf 42}, 78 (1990).

\bibitem{Cslmass}
Ghirardi, G.C. , Grassi, R. and Benatti, F. Describing the macroscopic world: closing the circle within the dynamical reduction program. {\it Found. Phys.} {\bf 25}, 5-38 (1995).

\bibitem{Di1}
Di\'osi, L. Quantum stochastic processes as models for state vector reduction. {\it J. Phys. A: Math. Gen.} {\bf 21}, 2885-2898 (1988).
\bibitem{Di2}
Di\'osi, L. Continuous quantum measurement and Ito formalism. {\it Phys. Lett. A} {\bf 129}, 419-423 (1988).

\bibitem{qmupl}
Di\'osi, L. Models for universal reduction of macroscopic quantum fluctuations. {\it Phys. Rev. A} \textbf{40}, 1165-1174 (1989).

\bibitem{Pd}
Penrose, P. On gravity's role in quantum state reduction. {\it Gen. Rel. Grav.} {\bf 28}, 581-600 (1996).


\bibitem{Pr}
Bassi, A. and Ghirardi, C. G. Dynamical reduction models. {\it Phys. Rep.} {\bf 379}, 257-426 (2003).

\bibitem{collapse_review2}
Bassi, A., Lochan, K., Satin, S., Singh, T. P. and Ulbricht, H. Models of wave-function collapse, underlying theories, and experimental tests. {\it Rev. Mod. Phys.} {\bf 85}, 471-527 (2013).

\bibitem{new_phys1}
Adler, S. L. {\it Quantum Theory as an Emergent Phenomenon}, Cambridge University Press (2004).
\bibitem{new_phys2} Adler, S. L., Bassi, A. Is quantum theory exact?. {\it Science} {\bf 325}, 275-276 (2009).
\bibitem{new_phys3} Weinberg, S. Collapse of the state vector. {\it Phys. Rev. A} \textbf{85}, 062116-21 (2012).

\bibitem{adlerphoto}
Adler, S. L. Lower and upper bounds on CSL parameters from latent image formation and IGM heating. {\it J. Phys. A} {\bf 40}, 2935-2958 (2007).

\bibitem{macro_mol1}
Arndt, M., Nairz, O., Vos-Andreae, J., Keller, C., van der Zouw, G., \& Zeilinger, A. Wave–particle duality of C60 molecules. {\it Nature} {\bf 401}, 680-682 (1999).
\bibitem{macro_mol2}
Gerlich, S. \textit{et al.} Quantum interference of large organic molecules. {\it Nature Comm.} \textbf{2}, 263-267 (2011).
\bibitem{macro_mol3}
Hornberger, K., Gerlich, S., Haslinger, P., Nimmrichter, S., and Arndt, M. Colloquium: Quantum interference of clusters and molecules. {\it Rev. Mod. Phys.} \textbf{84}, 157-173 (2012).


\bibitem{opto1}
Marshall, W., Simon, C., Penrose, R., Bouwmeester, D. Towards quantum superpositions of a mirror. {\it Phys. Rev. Lett.} \textbf{91}, 130401-4 (2003).
\bibitem{opto2} Romero-Isart, O. Quantum superposition of massive objects and collapse models. {\it Phys. Rev. A} \textbf{84}, 052121-37 (2011).
\bibitem{opto3} Aspelmeyer, M., Meystre, P., Schwab, K. Quantum Optomechanics. {\it Physics Today} \textbf{65} (7), 29-35 (2012).


\bibitem{bi}
Bassi, A. and Ippoliti, E. Numerical analysis of a spontaneous collapse model for a two-level system. {\it Phys. Rev. A} {\bf 69}, 012105-13 (2004).


\bibitem{Ch}
Christian, J. Testing gravity-driven collapse of the wave function via cosmogenic neutrinos. {\it Phys. Rev. Lett.} {\bf 95}, 160403-6 (2005).

\bibitem{neuCSL}
See the supplementary material on arXiv:1207.5997 (2012).

\bibitem{ggr}
Ghirardi, G. C., Grassi, R., Rimini, A.  Continuous-spontaneous-reduction model involving gravity. {\it Phys. Rev. A} \textbf{42}, 1057-1067 (1990).


\bibitem{sigmanue1}
Flowers, E. G., Sutherland, P. G. Neutrino-neutrino scattering and supernovae. {\it Astrophys. J.} {\bf 208}, L19-L21 (1976).
\bibitem{sigmanue2}
Marciano, W. J., Parsa, Z. Neutrino–electron scattering theory. {\it J. Phys. G} {\bf 29}, 2629-2645 (2003).

\bibitem{decoexp}
Fogli, G. L., {\it et al.} Observables sensitive to absolute neutrino masses: A reappraisal after WMAP 3-year and first MINOS results. {\it Phys. Rev. D} {\bf 76}, 033006-16 (2007).

\bibitem{kaonCSL}
See the supplementary material on arXiv:1207.6000 (2012).

\bibitem{BeatrixZeta3}
Bertlmann, R. A., Grimus, W., and Hiesmayr, B.C. Open-quantum-system formulation of particle decay. {\it Phys. Rev. A} \textbf{73}, 054101-4 (2006).

\bibitem{Appollo}
Go, A., {\it et al.} the Belle Collaboration. Measurement of Einstein-Podolsky-Rosen-Type Flavor Entanglement in Y($4S$)$\rightarrow B^{0}\bar{B}^0$ Decays. {\it Phys. Rev. Lett.} \textbf{99}, 131802-7 (2007).

\bibitem{CPLEAR}
Apostolakis, A., {\it et al.} CPLEAR collaboration, An EPR experiment testing the non-separability of the $K^0\bar{K}^0$ wave function. {\it Phys. Lett. B} \textbf{422}, 339-348 (1998).

\bibitem{DiDomenico1}
Ambrosino, F., {\it et al.} KLOE collaboration, First observation of quantum interference in the process $\phi\rightarrow K_S K_L\rightarrow\pi^+\pi^-\pi^+\pi^-$: A test of quantum mechanics and CPT symmetry. {\it Phys. Lett. B} {\bf 642}, 315 (2006).
\bibitem{DiDomenico2} Di Domenico, A., {\it et al.} KLOE collaboration, $CPT$ Symmetry and Quantum Mechanics Tests in the Neutral Kaon System at KLOE. {\it Found. Phys.} {\bf 40},  852-866 (2010).
\bibitem{DiDomenico3} Amelino-Camelia, G., {\it et. al.} KLOE collaboration, Physics with the KLOE-2 experiment at the upgraded DA$\Phi$NE. {\it Eur. Phys. J. C} {\bf 68}, 619-681 (2010).

\bibitem{ParticleDataBook}
Nakamura, K., \emph{et al.} (Particle Data Group), Review of Particle Physics. {\it J. Phys. G} \textbf{37}, 075021 (2010).

\bibitem{resol1}
Darquie, B. {\it et al.} Progress toward the first observation of parity violation in chiral molecules by high-resolution laser spectroscopy. {\it Chirality.} {\bf 22}, 870-884 (2010).
\bibitem{resol2}
Quack, M., Stohner, J. and Willeke, M. High-resolution spectroscopic studies and theory of parity violation in chiral molecules. {\it Annu. Rev. Phys. Chem.} {\bf 59}, 741-769 (2008).


\bibitem{so}
Fern\'{a}ndez, I. and Khiar, N.  Recent developments in the synthesis and utilization of chiral sulfoxides. {\it Chem. Rev.} \textbf{103} (9), 3651-705 (2003).

\bibitem{firefly}
Granovsky, Alex A. Firefly version 7.1.G, www
http://classic.chem.msu.su/gran/firefly/index.html (Date of access: June 2012).
Schmidt, M. W.  \emph{et al.}, General atomic and molecular electronic structure system. {\it J. Comput. Chem.} \textbf{14}, 1347-1362 (1993).


\bibitem{vacuum}
Gabrielse, G. \emph{et al.} Thousandfold improvement in the measured antiproton mass. {\it Phys. Rev. Lett.} \textbf{65}, 1317-1320 (1990).



\bibitem{dimer}
Tautermann, C. S., Voegele, A. F. and Liedl, K. R. The ground-state tunnelling splitting of various carboxylic acid dimers. {\it J. Chem. Phys.} {\bf 120}, 631-637 (2004).

\bibitem{tunnel2}
Buntkowsky, G., Limbach, H. H. H-solid state NMR studies of tunnelling
 phenomena and isotope effects in transition metal dihydrides. {\it J. Low. Temp. Phys.} 143, 55-114 (2006).

\end{thebibliography}
\end{document}